\documentclass[floatfix,aps,prl,twocolumn,showpacs,superscriptaddress]{revtex4-1}

\usepackage[english]{babel}
\usepackage{bm}    

\usepackage{epsfig}
\usepackage{amsbsy,amssymb} 
\usepackage{graphicx}
\usepackage{array}
\usepackage{amsmath}
\usepackage{multirow}

\usepackage{natbib}

\allowdisplaybreaks

\newcommand\new{\newcommand}         

\def\beq{\begin{equation}}   
\def\eeq{\end{equation}}
\def\bea{\begin{eqnarray}}  
\def\eea{\end{eqnarray}} 
\newcommand{\bite}{\begin{itemize}}
\newcommand{\eite}{\end{itemize}}

\def\nn{\nonumber}
\def\NN{\mathcal{N}}

\def\z{\mathbf{z}}
\def\G{\mathcal{G}}

\def\la{\langle}
\def\ra{\rangle}

\def\nn{\nonumber \\}

\newcommand\polia[2]{ \left\lfloor #1 \right\rfloor _{#2} }

\newcommand\GG[1]{\G_{#1}}

\newcommand\DK[2]{ D_{#1} }

\new{\eV}         {{\ifmmode {\mathrm{ eV}}\else ${\mathrm{ eV}}$\fi}}
\new{\MeV}        {{\ifmmode {\mathrm{ MeV}}\else ${\mathrm{ MeV}}$\fi}}
\new{\MeVc}       {{\ifmmode {\mathrm{ MeV}}/c\else ${\mathrm{ MeV}}/c$\fi}}
\new{\MeVcc}      {{\ifmmode {\mathrm{ MeV}}/c^2\else ${\mathrm{ MeV}}/c^2$\fi}}
\new{\GeV}        {{\ifmmode {\mathrm{ GeV}}\else ${\mathrm{ GeV}}$\fi}}
\new{\GeVc}       {{\ifmmode {\mathrm{ GeV}}/c\else ${\mathrm{GeV}}/c$\fi}}
\new{\GeVcc}      {{\ifmmode {\mathrm{ GeV}}/c^2\else ${\mathrm{GeV}}/c^2$\fi}}
\new{\TeV}        {{\ifmmode {\mathrm{ TeV}}\else ${\mathrm{ TeV}}$\fi}}

\new{\Mh}         {{\ifmmode M_{\mathrm{ H}}
                    \else $M_{\mathrm{H}}$\fi}}

\new{\Mz}         {{\ifmmode M_{\mathrm{Z}}
                    \else $M_{\mathrm{Z}}$\fi}}
\new{\Mzsq}       {{\ifmmode M^2_{\mathrm{ Z}}
                    \else $M^2_{\mathrm{Z}}$\fi}}

\new{\as}[1]      {{\ifmmode\alpha^{#1}_s
                    \else$\alpha^{#1}_s$\fi}}
\new{\asx}[1]      {{\ifmmode a^{#1}_s
                    \else $a^{#1}_s$\fi}}
\new{\asb}[1]     {{\ifmmode\overline{\alpha}^{#1}_s
                    \else $\overline{\alpha}^{#1}_s$\fi}}
\new{\asmz}       {{\ifmmode\alpha_s(\Mzsq)
                    \else $\alpha_s(\Mzsq)$\fi}}
\new{\lqcd}       {{\ifmmode\Lambda_{\mathrm{ QCD}}
                    \else $\Lambda_{\mathrm{ QCD}}$\fi}}

\begin{document}

\title{Multiloop Integrand Reduction for Dimensionally Regulated Amplitudes}

\author{P.~Mastrolia}
\affiliation{Max-Planck-Institut f\"ur Physik, F\"ohringer Ring 6,
80805 M\"unchen, Germany}
\affiliation{Dipartimento di Fisica e Astronomia, Universit\`a di
Padova, and INFN
Sezione di Padova, via Marzolo 8, 35131 Padova, Italy}
\author{E.~Mirabella}
\affiliation{Max-Planck-Institut f\"ur Physik, F\"ohringer Ring 6,
80805 M\"unchen, Germany}
\author{G.~Ossola}
\affiliation{New York City College of Technology, City University of New York, 300 Jay Street, Brooklyn NY 11201, USA}
\affiliation{The Graduate School and University Center, City University of New York, 365 Fifth Avenue, New York, NY 10016, USA}
\author{T.~Peraro}
\affiliation{Max-Planck-Institut f\"ur Physik, F\"ohringer Ring 6,
80805 M\"unchen, Germany}


\begin{abstract}
We present the integrand reduction via multivariate polynomial
division as a natural technique to encode the unitarity conditions
of Feynman amplitudes.  We derive a recursive formula for the
integrand reduction, valid for arbitrary dimensionally regulated loop
integrals with any number of loops and external legs, which can be
used to obtain the decomposition of any integrand analytically with a
finite number of algebraic operations.  The general results are
illustrated by applications to two-loop Feynman diagrams in QED and
QCD, showing that the proposed reduction algorithm can also be seamlessly
applied to integrands with denominators appearing with arbitrary
powers.
\end{abstract}

 

\maketitle

%
%
%

\paragraph{Introduction --}
In the perturbative approach to quantum field theories, the elements
of the scattering matrix, which are  the scattering amplitudes, can be
expressed in terms of Feynman diagrams. The latter  generally represent
multiple integrals whose integrand is a rational function of the
integration variables.  Scattering amplitudes are analytic functions
of the kinematic variables of the interacting particles, hence they
are determined by their singularities, whose
location in the complex plane is specified by a set of algebraic
equations.  The analysis of the singularity structure 
can be used to define the discontinuities
of a Feynman integral across the branch cuts
attached to the Landau singularities. They  are encoded in the 
Cutkosky formula and  correspond to the unitarity conditions of the scattering amplitude.  
In the canonical formalism, the unitarity cut conditions have been used for the 
evaluation of the scattering amplitudes trough dispersive Cauchy's integral
representations.   However,
the dispersive approach is well-known to suffer from
ambiguities which limit its applicability for the quantitative
evaluation of generic Feynman integrals in gauge theories.

In the more modern interpretation of unitarity,  cut conditions and
analyticity are   successfully exploited for decomposing
scattering amplitudes in terms of independent functions -- rather than
for  their  direct evaluation.  The basic functions entering the
amplitudes decomposition are univocally characterized by their
singularities.
The singularity structure can be accessed before integration, at the
integr{\it and} level \cite{Ossola:2006us,Mastrolia:2011pr}.
Therefore, the decomposition of the integrated amplitudes can be
deduced from the the decomposition of the corresponding 
integr{\it ands}.
The integrand-reduction
methods~\cite{Ossola:2006us,Ellis:2007br,Mastrolia:2011pr,Badger:2012dp,Mastrolia:2012bu,Zhang:2012ce,Mastrolia:2012an}
rely on the existence of a relation between the numerator and the
denominators of each Feynman integral. A generic numerator can be
expressed as a combination of (products of) denominators, multiplied
by polynomial coefficients, which correspond to the {\it residues} at
the multiple cuts of the diagrams. The multiple-cut conditions,
generally fulfilled for complex values of the integration variables,
can be viewed as {\it projectors} isolating each residue. 
The latter,
depicted as an on-shell cut diagram, represents the amplitude factorized
into a product of simpler amplitudes, either with fewer loops or a lower number of
legs.

\begin{figure}[t!]
\includegraphics[width=8.3cm]{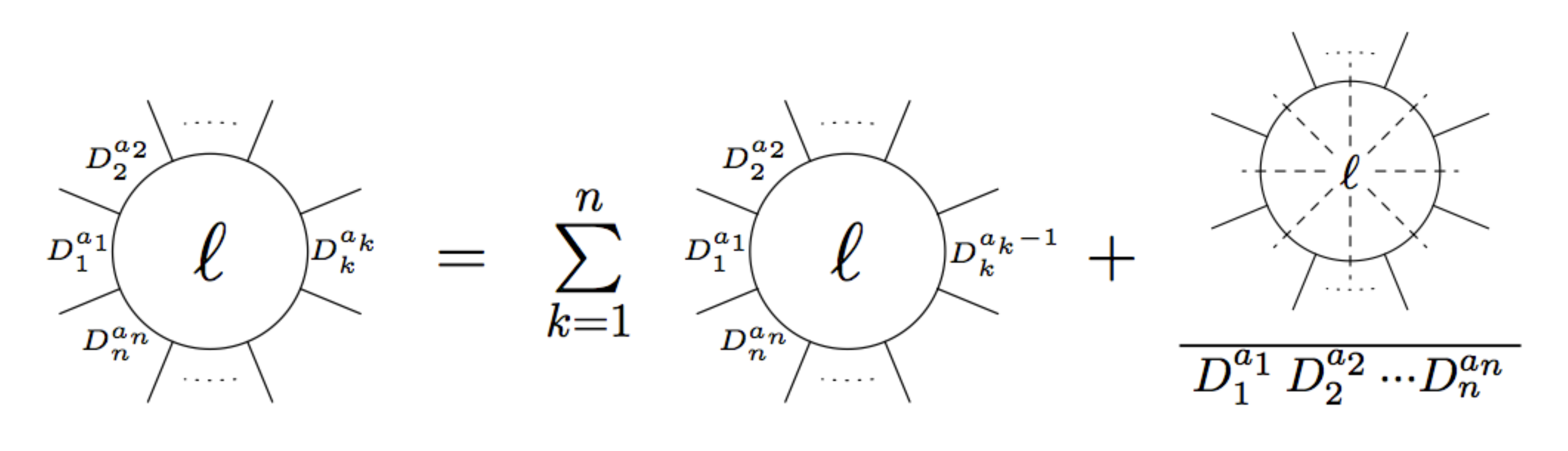}
\caption{Integrand recurrence relation for a generic $\ell$-loop integrand.} \label{Fig:formula}
\end{figure}

The residues are multivariate polynomials in those
components of the propagating momenta which 
correspond to {\it irreducible scalar products} (ISPs), 
that cannot be decomposed in terms of denominators. The ISPs
either yield {\it spurious} contributions, which 
vanish upon integration, or 
generate the basic integrals entering the 
amplitude decomposition~\cite{Mastrolia:2011pr,Badger:2012dp}.

Within the integrand reduction methods, the problem of decomposing any 
scattering amplitude in terms of independent integrals is therefore reduced 
to the algebraic problem of reconstructing the residues at its multiple cuts.

In Refs.~\cite{Zhang:2012ce,Mastrolia:2012an} the determination of the
residues at the multiple cuts has been formulated as a problem of {\it
multivariate polynomial division}, and solved using algebraic geometry 
techniques. These techniques  allowed  
one to prove that the integrand
decomposition, originally formulated for one-loop amplitudes~\cite{Ossola:2006us}, is
valid and applicable at any order in perturbation theory, 
irrespective of the complexity of the topology of the involved diagrams, 
being them massless or massive, planar or non-planar. 
This novel reduction algorithm 
has been applied to the decomposition of supersymmetric amplitudes at two and 
three loops~\cite{Badger:2012dv,Mastrolia:2012wf}.
Also, it has been used for the identification of the two-loop integrand 
basis in four dimensions~\cite{Feng:2012bm}, and for the  classification of the
cut solutions according to  the geometric properties of the associated 
varieties~\cite{CaronHuot:2012ab,Huang:2013kh}.

In Ref.~\cite{Mastrolia:2012an}, we found an 
{\it integrand-recursion formula} for the iterative decomposition of scattering 
amplitudes, based on successive divisions of 
the numerators modulo the Gr\"obner basis of the ideals generated 
by the cut denominators.
The integrand recurrence relation may be applied in two ways. 

The first approach, that we define {\it fit-on-the-cuts}, 
requires the  knowledge of the parametric residues and of the parametric 
(families of) solutions of all possible multiple cuts. The parameters of the residues 
are determined by evaluating the numerator
at the solutions of the multiple cuts,  as many times as the number of the unknown coefficients. 
This approach is  the canonical way to achieve the integrand
decomposition of scattering amplitudes at one loop
\cite{Ossola:2006us}, and it has been implemented in public codes like  
{\sc Cutools}~\cite{Ossola:2007ax}, 
and {\sc Samurai}~\cite{Mastrolia:2010nb}.
In this approach the parametrization of the residues can be found 
 by applying  the  integrand-recursion formula to 
 the most generic numerator function, with parametric coefficients.
 
Alternatively, as we show in this letter, the reduction
formula can be applied directly to the numerator, 
within what we define as the  {\it divide-and-conquer} approach.
In this case, the decomposition of the amplitude is obtained by successive
polynomial divisions, which at each step  generate the actual residues. 
In this way, 
the decomposition of any integrand is obtained analytically, with a finite
number of algebraic operations,
without requiring the knowledge of the varieties of solutions of the multiple cuts, nor the one of the parametric form of the residues.

In the following, we describe the coherent mathematical framework underlying
the integrand decomposition, interpreting the 
unitarity-cut conditions as {\it equivalence classes} of polynomials.
We present the {\it divide-and-conquer} approach through its systematic  
application to the decomposition of some classes of two-loop diagrams. 
The  examples show the main features of the proposed 
reduction algorithm,  which can 
be applied  to generic dimensionally regulated Feynman integrals with 
multiple denominators, namely  denominators appearing with arbitrary powers.
To the best of our knowledge, this is the  first application of 
integrand-reduction algorithms directly to diagrams  with 
multiple propagators. 

With this communication  we finally  aim at presenting the
integrand reduction via multivariate polynomial division as a
natural technique to encode  the unitarity conditions of Feynman amplitudes.
Indeed Cauchy's integration, which is the underlying concept 
of unitarity integrals and, more generally, of discontinuities formulas, 
when applied to rational integrands corresponds to partial fraction, which
is the  objective of the polynomial division.
\medskip


\paragraph{Integrand reduction formula --} The extension 
of the integrand recurrence relation required to accommodate 
multiple propagators  is straightforward.  
An arbitrary graph with $\ell$ loops represents a $d$-dimensional  integral  of the type
\begin{align}
& \int \, d^d \bar q_1 \cdots  d^d \bar q_\ell  \; \mathcal{I}_{
{\tiny \underbrace{i_1\cdots i_1}_{a_1} } \, {\tiny \ldots} \,  {\tiny \underbrace{i_n\cdots i_n}_{a_n}} 
}  \, ,\nn
& \mathcal{I}_{
{\tiny \underbrace{i_1\cdots i_1}_{a_1} } \, {\tiny \ldots} \,  {\tiny \underbrace{i_n\cdots i_n}_{a_n}} 
}
\equiv \frac{\mathcal{N}_{i_1 \cdots i_1 \, \cdots \, i_n \cdots i_n}
 }{D^{a_1}_{i_1}\cdots D^{a_n}_{i_n}} \, ,
\label{Eq:integrand}
\end{align}
with $i_1,\ldots ,i_n$ distinct indices. The numerator and the denominators are
polynomials in a set of coordinates $\z$, {\it i.e.}\ they are in the polynomial ring
$P[\z]$. We define the ideal 
\begin{align*}
\mathcal{J}_{{\tiny \underbrace{i_1\cdots i_1}_{a_1} }  \, {\tiny \ldots} \,  {\tiny \underbrace{i_n\cdots i_n}_{a_n}} }
& \equiv  \la \underbrace{ D_{i_1}, \ldots , D_{i_1}}_{a_1}, \ldots,  \underbrace{D_{i_n}, \ldots , D_{i_n}}_{a_n} \ra \, ,
\end{align*}
which fulfills the relation
\begin{align}
 \mathcal{J}_{{\tiny \underbrace{i_1\cdots i_1}_{a_1} } \,  {\tiny \underbrace{i_2\cdots i_2}_{a_2} } \, {\tiny \ldots} \,  {\tiny \underbrace{i_n\cdots i_n}_{a_n}} } & =
\mathcal{J}_{i_1i_2\cdots i_n} =\,   \nn
=   \Bigg \{  \sum_{k=1}^n \,  h_k(\z)  \;& 
  D_{i_k}(\z) \; :  \; h_k(\z) \in P[\z]   \, 
\Bigg \} \, .
\label{Eq:JEJ}
\end{align}
Given a monomial ordering, we define the {\it normal form}  of  
a polynomial $p(\z)$ with respect to the ideal $\mathcal{J}$ as
\begin{align} 
\polia{p(\z)}{\mathcal{J}_{i_1\cdots i_n}     }  \equiv  p(\z) ~ \textrm{mod} ~ \GG{i_1 \cdots i_n}  \, ,
\end{align}  
{\it i.e.}\ the normal form of $p$ is the remainder of its  division modulo a  Gr\"obner  basis  $\G$ of 
 $\mathcal{J}$. Two polynomials $p(\z), q(\z) \, \in P[\z]$ are {\it congruent modulo}
 $\mathcal{J}$ iff their difference can be written in terms of the denominators, {\it i.e.}
\begin{align*} 
p(\z)\sim_{\mathcal{J}_{i_1\cdots i_n}} q(\z) \quad  \mbox{ iff }
\quad  
p(\z) - q(\z)  \in \mathcal{J}_{i_1\cdots i_n}\; .
\end{align*}
The congruence modulo $\mathcal{J}$
is an equivalence relation  and the set 
of all its  equivalence
classes is the {\it quotient ring} 
$P[\z] / \mathcal{J}$.  The properties of the Gr\"obner basis
ensure  that
\begin{align*} 
p(\z)\sim_{\mathcal{J}_{i_1\cdots i_n}} q(\z)  \quad  \mbox{ iff }
\quad  
\polia{p(\z)}{\mathcal{J}_{i_1\cdots i_n}     } =
\polia{q(\z)}{\mathcal{J}_{i_1\cdots i_n}     } \, .
\end{align*}  
Therefore, the normal form of the elements of the equivalence classes
establish a natural correspondence between  $P[\z] /\mathcal{J}$  and  $P[\z]$.

The numerator $\mathcal{N}$ of Eq.~(\ref{Eq:integrand}) is a polynomial in $\z$
and can be decomposed by performing  the division 
\begin{align}
\mathcal{N}_{
{\tiny \underbrace{i_1\cdots i_1}_{a_1} } \, {\tiny \ldots} \,  {\tiny \underbrace{i_n\cdots i_n}_{a_n}} 
} \, 
/  \,
\GG{
{\tiny \underbrace{i_1\cdots i_1}_{a_1} } \, {\tiny \ldots} \,  {\tiny \underbrace{i_n\cdots i_n}_{a_n}} 
} \, .
\end{align}
Eq.~(\ref{Eq:JEJ}) allows one to write its  decomposition as
\begin{align}
\mathcal{N}_{i_1 \cdots i_1 \, \cdots \, i_n \cdots i_n} ={}& 
\Gamma_{i_1 \cdots i_1 \, \cdots \, i_n \cdots i_n}  + \nn
+{}& \polia{\mathcal{N}_{i_1 \cdots i_1 \, \cdots \, i_n \cdots i_n}
  }{ \mathcal{J}_{i_1i_2\cdots i_n}  } \, .
\label{Eq:Enne}
\end{align}

The normal form of the numerator is not in the ideal $\mathcal{J}$, thus it cannot be expressed in terms of the denominators  and 
it is  identified with the residue of the multiple cut $D^{a_1}_{i_1} = \cdots = D^{a_n}_{i_n} =0$,
\begin{equation}
  \polia{\mathcal{N}_ {i_1 \cdots i_1 \, \cdots \, i_n \cdots i_n}  }{ \mathcal{J}_{i_1i_2\cdots i_n}  } = \Delta_ {i_1 \cdots i_1 \, \cdots \, i_n \cdots i_n}  \,  ,
 \label{Eq:Delta}
\end{equation}
belonging to  the quotient ring $P[\z] / \mathcal{J}$.
The term  $\Gamma$, instead,  belongs to the ideal $\mathcal{J}$, thus it can be written  as 
\begin{eqnarray}
\Gamma_ {i_1 \cdots i_1 \, \cdots \, i_n \cdots i_n}   &=& \sum_{k=1}^n \mathcal{N}_{{\tiny \underbrace{i_1\cdots i_1}_{a_1} } \, {\tiny \ldots} \,  {\tiny \underbrace{i_k\cdots i_k}_{a_k-1}}
 {\tiny \ldots} \,  {\tiny \underbrace{i_n\cdots i_n}_{a_n}}}
  D_{i_k}  \, .
\label{Eq:Gamma}
\end{eqnarray}
Substituting  Eqs.~(\ref{Eq:Enne}), (\ref{Eq:Delta}), and~(\ref{Eq:Gamma}) in  Eq.~(\ref{Eq:integrand}), we obtain 
\begin{align}
\mathcal{I}_{{\tiny \underbrace{i_1\cdots i_1}_{a_1} } \, {\tiny \ldots} \,   {\tiny \underbrace{i_n\cdots i_n}_{a_n}}}  ={}&
  \sum_{k=1}^n \,  \mathcal{I}_{{\tiny \underbrace{i_1\cdots i_1}_{a_1} } \, {\tiny \ldots} \,  {\tiny \underbrace{i_k\cdots i_k}_{a_k-1}}
 {\tiny \ldots} \,  {\tiny \underbrace{i_n\cdots i_n}_{a_n}}} +  \nn
+{}&
 \frac{\Delta_{i_1\cdots i_1 \, \cdots \, i_n\cdots i_n}  }{ D^{a_1}_{i_1}\cdots D^{a_k}_{i_n} } \, ,
 \label{Eq:Recurrence}
\end{align}
which  is a  non-homogeneous 
recurrence relation expressing a given  integrand in terms of
integrands  with fewer denominators. it  is the generalization of the 
recurrence relation of Ref.~\cite{Mastrolia:2012an},  valid
for arbitrary powers of the denominators.
Its pictorial representation  is shown in Figure~\ref{Fig:formula}.
Within the {\it divide-and-conquer} approach,  the integrand reduction formula becomes  an elegant and  powerful tool
to perform the analytic decomposition of multi-loop integrals through a top-down procedure starting from the
integrand with the highest number of denominators. 
It is worth noticing that, in this algorithm, the presence of multiple denominators is reflected by the fact that the 
division modulo the ideal $\mathcal{J}_{i_1\cdots i_n}$ enters the procedure $a_1\times\cdots\times a_n$ times.
\medskip

In the following we apply  the {\it divide-and-conquer} approach to some  two- and three-point  two-loop
diagrams  appearing in  QED and  QCD radiative corrections. The  divergences have been  regularized 
within the 't Hooft--Veltman  scheme and the  computation has been carried out in the Feynman  gauge.
The decompositions  have  been verified by using the $N=N$ global test~\cite{Ossola:2006us, Ossola:2007ax, Mastrolia:2010nb}.
\medskip

\begin{figure}[t!]
\includegraphics[height=3.2cm]{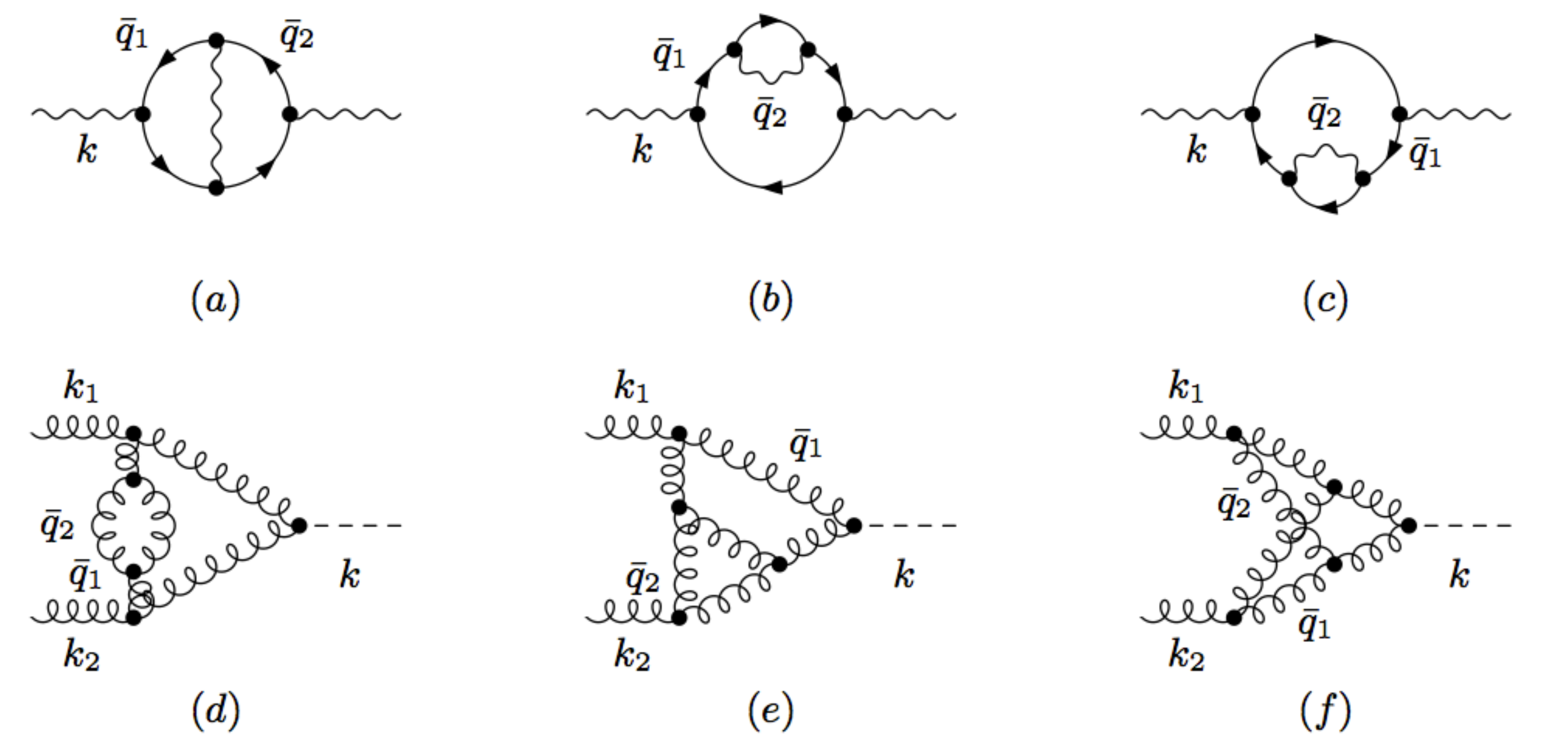}
\caption{First row:  diagrams  leading to the two-loop  QED corrections to the photon self energy.
Second row: two-loop diagrams entering the QCD corrections to $gg \to H$ in the heavy top mass approximation.
} \label{Fig:diags}
\end{figure}

\paragraph{Photon vacuum polarization --}
As a first example we consider  the two-loop  contributions
to the transverse part $\Pi(k^2)$  of the  vacuum polarization in QED with a massive fermion~\cite{Broadhurst:1993mw}.
The integrand of $\Pi(k^2)$ 
gets contributions  from the three 
self-energy diagrams in the first row of  Figure~\ref{Fig:diags}.  
The $d$-dimensional loop momenta $\bar q_{i}$
are split into a  $4$-dimensional and $(-2\,\epsilon)$-dimensional part, $\bar q_i = q_i +\vec \mu_i$, with
$q_i\cdot \vec \mu_j=0$ and $\vec \mu_i\cdot \vec \mu_j \equiv \mu^2_{ij}$. In this case the 
variables $\z$ are  $\mu^2_{11}$, $\mu^2_{22}$, $\mu^2_{12}$ and  
the components of $q_i$  in the basis $\{k, k_\perp, e_3, e_4\}$, such that 
\begin{align*}
k_\perp^2 \neq 0 \neq e_3\cdot e_4\, , \quad
k\cdot k_\perp =k\cdot e_{j} =  k_\perp \cdot e_{j} = e_{j}^2 =0 \, .
\end{align*}

The integrand of the diagram $(a)$ is  
\begin{align}
 \mathcal{I}^{(a)}_{12345}=  &\,  \frac{1}{3-2\, \epsilon} \frac{ \mathcal{N}^{(a)}_{12345}}{D_1D_2D_3D_4 D_5} \, , 
\end{align}
while its denominators are
\begin{align*}
&\DK{1}{a} =\bar q_1^2 -m^2  \, , && \DK{2}{a} =\left (\bar q_1+k\right )^2 -m^2  \, , \nn
&\DK{3}{a} =\bar q_2^2 -m^2 \, ,  &&  \DK{4}{a} =\left (\bar q_2+k\right )^2 -m^2  \, ,\nn
&\DK{5}{a} = (\bar q_1- \bar q_2)^2 \, .&&  
\end{align*}
According to our algorithm,  the first step of the reduction requires the division
$\NN^{(a)}_{12345} / \GG{12345}$, whose result reads as
\begin{align}
 \mathcal{N}^{(a)}_{12345} =&\, 
\Delta_{12345} +\NN_{1235}\DK{4}{a} +\mathcal{N}_{2345}\DK{1}{a} +\mathcal{N}_{1345}\DK{2}{a}  \nn
 +& \mathcal{N}_{1245}\DK{3}{a}    +\mathcal{N}_{1234}\DK{5}{a}   \, .
 \label{Eq:NCS1}
 \end{align}
 In the second step, the numerators  $\mathcal{N}_{i_1i_2i_3i_4}$   are reduced performing the division $\mathcal{N}_{i_1i_2i_3i_4} /  \GG{i_1i_2i_3i_4}$,
 \begin{align}
 \mathcal{N}^{(a)}_{12345} =&\, \Delta_{12345} +\Delta_{1235}\DK{4}{a} +\Delta_{2345}\DK{1}{a} +\Delta_{1345}\DK{2}{a}  \nn
 +& \Delta_{1245}\DK{3}{a}    +\Delta_{1234}\DK{5}{a}  + \NN_{123}\DK{4}{a} \DK{5}{a} \nn
 +& \NN_{124}\DK{3}{a} \DK{5}{a} + \NN_{134}\DK{2}{a} \DK{5}{a}+ \NN_{234}\DK{1}{a} \DK{5}{a} \nn
 +& \NN_{125}\DK{3}{a} \DK{4}{a} + \NN_{135}\DK{2}{a} \DK{4}{a}  + \NN_{245}\DK{1}{a} \DK{3}{a} \nn
 +& \NN_{345}\DK{1}{a} \DK{2}{a} + \NN_{145}\DK{2}{a} \DK{3}{a} + \NN_{235}\DK{1}{a} \DK{4}{a}   \, .
 \label{Eq:NCS2}
 \end{align}
 The complete decomposition of $\NN^{(a)}_{12345}$ is obtained by iterating the procedure twice,
\begin{align}
 \mathcal{N}^{(a)}_{12345} =&\, 
\Delta_{12345} +\Delta_{1235}\DK{4}{a} +\Delta_{2345}\DK{1}{a} +\Delta_{1345}\DK{2}{a}  \nn
 +& \Delta_{1245}\DK{3}{a}    +\Delta_{1234}\DK{5}{a}  + \Delta_{123}\DK{4}{a} \DK{5}{a} \nn
 +& \Delta_{124}\DK{3}{a} \DK{5}{a} + \Delta_{134}\DK{2}{a} \DK{5}{a}+ \Delta_{234}\DK{1}{a} \DK{5}{a} \nn
 +& \Delta_{125}\DK{3}{a} \DK{4}{a} + \Delta_{135}\DK{2}{a} \DK{4}{a}  + \Delta_{245}\DK{1}{a} \DK{3}{a} \nn
 +& \Delta_{345}\DK{1}{a} \DK{2}{a} + \Delta_{145}\DK{2}{a} \DK{3}{a} + \Delta_{235}\DK{1}{a} \DK{4}{a} \nn
 +& \Delta_{13}\DK{2}{a} \DK{4}{a} \DK{5}{a} +\Delta_{24}\DK{1}{a} \DK{3}{a} \DK{5}{a}  \nn
 +& \Delta_{14}\DK{2}{a} \DK{3}{a} \DK{5}{a} +\Delta_{23}\DK{1}{a} \DK{4}{a} \DK{5}{a}  \, .
  \label{Eq:Pia}
\end{align}
The residues in  Eq.~(\ref{Eq:Pia}) read as follows:
\begin{align}
 \Delta_{12345}&= 8\left ( 4\, m^4 - k^4  +  k^2\,( k^2 -2\,m^2) \, \epsilon  \right)\, , \nn
\Delta_{1234}&= -4 \, \bigg [\, \big(  4\, m^2 +k^2( 3  -  \epsilon -  2\, \epsilon^2) \big ) \nn
   &\quad +  4\, \left (1-\epsilon\right )\, \bigg ( \mu^2_{12} - \frac{(q_1 \cdot k_\perp)\, (q_2 \cdot k_\perp)}{k_\perp^2}  \nn
   &\quad   - \frac{ (q_1 \cdot e_3)\, (q_2 \cdot e_4)}{ (e_3 \cdot e_4)}   -  \frac{(q_1 \cdot e_4)\, (q_2 \cdot e_3)}{(e_3 \cdot e_4)}   \bigg )  \bigg ] \, ,\nn
 \Delta_{1235}&= \Delta_{2345} = \Delta_{1345} = \Delta_{1245} =8\left (m^2 +k^2 (1- \epsilon) \right ) \, ,\nn
\Delta_{123}&= \Delta_{124} = \Delta_{134}= \Delta_{234} =4\left (1- \epsilon \right )\, ,\nn
 \Delta_{125}&= \Delta_{135}= \Delta_{245}= \Delta_{345} =-8\left (1- \epsilon \right )\, ,\nn
 \Delta_{145}&= \Delta_{235}=8\, \epsilon \, \left (1- \epsilon \right )\, ,\nn
  \Delta_{13}&= \Delta_{24}= -\Delta_{14} = -\Delta_{23} = \frac{4\left (1-\, \epsilon \right )}{k^2} \, .
 \end{align}
The diagram $(b)$ contains a double propagator, 
\begin{align}
 \mathcal{I}^{(b)}_{11234} =  &\,  \frac{1}{3-2\, \epsilon} \frac{ \mathcal{N}^{(b)}_{11234} }{D_1^2D_2D_3D_4} \, , 
\end{align}
where the  denominators are
\begin{align*}              
&\DK{1}{a} =\bar q_1^2 -m^2  \, , && \DK{2}{a} =\left (\bar q_1-k\right )^2 -m^2 \, , \nn
&\DK{3}{a} =\bar q_2^2 \, ,  &&  \DK{4}{a} =\left (\bar q_1+\bar q_2 \right )^2 -m^2  \, .
\end{align*}
The first step of the reduction requires the division $\NN^{(b)}_{11234} / \GG{11234}$ which, because of  Eq.~(\ref{Eq:JEJ}),  is 
equivalent to the division $\NN^{(b)}_{11234} / \GG{1234}$,  
\begin{align}
\mathcal{N}^{(b)}_{11234}  ={}&  \Delta_{11234}+  \mathcal{N}_{1234} \DK{1}{b} +   \mathcal{N}_{1123}\DK{4}{b}  \nn
+{}& \mathcal{N}_{1134}\DK{2}{b}  +\mathcal{N}_{1124}\DK{3}{b} \, .
\end{align}
In the  second step we perform  the divisions  $\mathcal{N}_{i_1 i_2 i_3 i_4} / \GG{i_1 i_2 i_3 i_4}$, obtaining
\begin{align}
  \mathcal{N}^{(b)}_{11234} =  &\,  \Delta_{11234}+  \Delta_{1234} \DK{1}{b} +   \Delta_{1123}\DK{4}{b}  +\Delta_{1134}\DK{2}{b}  \nn
  +&\Delta_{1124}\DK{3}{b}+  \mathcal{N}_{113}\DK{2}{b}\DK{4}{b} +    \mathcal{N}_{114}\DK{2}{b}\DK{3}{b}  \nn
  +&  \mathcal{N}_{234}\DK{1}{b}^2  \, .
\end{align}
Due to Eq.~(\ref{Eq:JEJ}), the  division $\mathcal{N}_{1 1  i_3 i_4} / \GG{11 i_3 i_4}$ is
equivalent to $\mathcal{N}_{1 1  i_3 i_4} / \GG{1 i_3 i_4}$. 
The reduction is completed by performing the divisions  $\mathcal{N}_{i_1  i_2 i_3} / \GG{i_1 i_2 i_3}$, along the lines of the 
previous steps, obtaining
\begin{align}
  \mathcal{N}^{(b)}_{11234} =  &\,  \Delta_{11234}+  \Delta_{1234} \DK{1}{b} +   \Delta_{1123}\DK{4}{b}  +\Delta_{1134}\DK{2}{b}  \nn
  +&\Delta_{1124}\DK{3}{b}+  \Delta_{113}\DK{2}{b}\DK{4}{b} +    \Delta_{114}\DK{2}{b}\DK{3}{b}  \nn
  +&  \Delta_{234}\DK{1}{b}^2  \, ,
\end{align}
in terms of the residues 
\begin{align}         
  \Delta_{11234} &= 16m^2\, \left ( k^2+2\, m^2 -k^2\epsilon \right ) \, ,\nn 
  \Delta_{1234} &= 16 \, \left [ (q_2\cdot k) (1-\epsilon)^2 +m^2\right ] \, ,\nn
 \Delta_{1124} &=-\Delta_{1123}  =8\, (1-\epsilon) \, \left [   k^2 (1-\epsilon) + 2\, m^2   \right] \, ,\nn
\Delta_{1134} &=-16 m^2\,\left (  1-  \epsilon   \right)\, , \nn
\Delta_{113} &=-\Delta_{114} =\Delta_{234} = 8\, \left (  1-  \epsilon   \right)^2  \, .
\label{Eq:ResA}
\end{align}
The integrand  of the diagram $(c)$  is  obtained by performing the 
replacement $\mathcal{I}^{(c)}_{11234} = \mathcal{I}^{(b)}_{11234 \; | \; k \, \to\, -k}$.

We remark that the residues  can also be  expressed  in terms of normal forms. For instance,  in the case of $\mathcal{I}^{(b)}$,  $\Delta_{1123}$ and $\Delta_{113}$ can be written as
\begin{align*}   
    \Delta_{1123} & =
    \frac{ 
     \polia{   \mathcal{N}^{(b)}_{11234} - \Delta_{11234}  }{\mathcal{J}_{123} }
     }{
     \polia{ \DK{4}{a}  }{\mathcal{J}_{123} } 
     }\, , \\
\Delta_{113} & =
    \frac{ 
     \polia{   \mathcal{N}^{(b)}_{11234} - \Delta_{11234} - \Delta_{1123}D_4 - \Delta_{1134}D_2 }{\mathcal{J}_{13} }
     }{
     \polia{ \DK{2}{a}  \DK{4}{a}  }{\mathcal{J}_{13} } 
     }\, . \nonumber
  \end{align*}
  Since  $\Delta_{1234} D_1 \in \mathcal{J}_{1234}$,  the residue $\Delta_{1234}$ can be obtained using
\begin{align*}         
    \Delta_{1234} & =
   \polia{
    \frac{ 
     \polia{   \mathcal{N}^{(b)}_{11234}- \Delta_{11234}  }{\mathcal{J}_{1^2234} }
     }{
     \polia{ \DK{1}{a}  }{\mathcal{J}_{1^2 234} }
     }
     }{ \mathcal{J}_{1234}   } \, ,
\end{align*}
where  $\mathcal{J}_{1^2234} \equiv \la  \DK{1}{a}^2, \DK{2}{a}, \DK{3}{a}, \DK{4}{a} \ra \subset \mathcal{J}_{1234}$.
\medskip

\paragraph{Diagrams for Higgs production via gluon fusion --}
We also  consider the three-point  diagrams in the second row of Figure~\ref{Fig:diags}, which enter the two-loop QCD corrections to the 
Higgs production via gluon fusion in the heavy top limit~\cite{Anastasiou:2002yz}.  In this case,  
the variables $\z$ are  $\mu^2_{11} $, $\mu^2_{22} $, $\mu^2_{12} $
and the components of  the four-vectors $q_i$ in the basis of massless vectors  $\{k_1, k_2, e_3, e_4\}$, such that
$k_i\cdot e_j  =0$ and $e_3\cdot e_4 \neq 0$.    
Within the {\it divide-and-conquer} approach, the integrand of the generic  diagram  is  decomposed as
\begin{align}       
\mathcal{I}^{(x)} = \sum_{\kappa=2}^6 \,   \sum_{\{i_1\cdots i_\kappa\}} \, \frac{\Delta_{i_1\cdots i_\kappa}}{D_{i_1}\cdots D_{i_\kappa}} \, , \qquad x=d,e,f \, .
\end{align}       
For the diagram $(d)$ the second sum runs over the unordered selections without repetition 
 of $\{1,1,2,3,4,5\}$, while for 
the diagram $(e)$ and  $(f)$ it runs over  the unordered selections without repetition  of   $\{1,\ldots, 6\}$.
The expression  of the residues are  lengthy and are omitted, however they are available upon request.  
\medskip

The reduction  algorithm described in this letter has been automated in a python package which uses  {\sc Macaulay2}~\cite{M2} and {\sc Form}~\cite{Kuipers:2012rf}.  The 
 numerators  of the presented  examples  have been generated with {\sc QGraf}~\cite{Nogueira:1991ex} and {\sc Form}  and
independently with {\sc FeynArts}~\cite{Hahn:2000kx},  {\sc FeynCalc}~\cite{Mertig:1990an}, and {\sc FormCalc}~\cite{Agrawal:2011tm}. 
\medskip

\begin{acknowledgments}
We thank  Simon Badger and Yang
Zhang for comments on the manuscript.
The work of P.M. and T.P. was supported by the Alexander von
Humboldt Foundation, in the framework of the Sofja Kovalevskaja Award Project
``Advanced Mathematical Methods for Particle Physics'', endowed by the German
Federal Ministry of Education and Research.  G.O. was
supported in part by the NFS Grant PHY-1068550.
\end{acknowledgments}


\bibliography{references.bib}
\end{document}